%% file: pecvel.tex
\documentclass[useAMS,usenatbib,manuscript]{mn2e}
\usepackage{rotating}       
\usepackage{graphicx}
\usepackage{amsmath} 
\usepackage{amssymb} 
\usepackage{natbib}
\usepackage{url}

\input{HEADER}

\title[]{Bayesian redshift-space distortions correction from galaxy redshift surveys}

\author[F.~S.~Kitaura et.~al.]{Francisco-Shu Kitaura$^{1}$\thanks{E-mail: kitaura@aip.de, Karl-Schwarzschild-fellow}, Metin Ata$^{1}$, Raul E.~Angulo$^2$,  Chia-Hsun Chuang$^{3}$,  \and Sergio Rodr{\'i}guez-Torres$^{3,4,6}$, Carlos Hern{\'a}ndez Monteagudo$^2$,  Francisco Prada$^{4,5,6}$,  \and \& Gustavo Yepes$^6$   \\
$^{1}$Leibniz-Institut f\"ur Astrophysik Potsdam (AIP), An der Sternwarte 16, D-14482 Potsdam, Germany\\
$^{2}$Centro de Estudios de F{\'i}sica del Cosmos de Arag\'on (CEFCA), Plaza San Juan, 1, planta 2, E-44001 Teruel, Spain,\\
$^{3}$Instituto de F{\'i}sica Te{\'o}rica, (UAM/CSIC), Universidad Aut{\'o}noma de Madrid, Cantoblanco, E-28049 Madrid, Spain\\
$^{4}$Campus of International Excellence UAM+CSIC, Cantoblanco, E-28049 Madrid, Spain \\
$^{5}$Instituto de Astrof{\'i}sica de Andaluc{\'i}a (CSIC), Glorieta de la Astronom{\'i}a, E-18080 Granada, Spain\\
$^{6}$Departamento de F{\'i}sica Te{\'o}rica,  Universidad Aut{\'o}noma de Madrid, Cantoblanco, 28049, Madrid, Spain\\
}

\date{\today}

\begin{document}

\maketitle

\begin{abstract}
We present a Bayesian reconstruction method which maps a galaxy distribution from redshift-space to real-space inferring the distances of the individual galaxies. The method is based on sampling density fields assuming a lognormal prior with a likelihood given by the negative binomial distribution function modelling stochastic bias. We assume a deterministic bias given by a power law  relating the dark matter density field to the expected halo or galaxy field. Coherent  redshift-space distortions are corrected in a Gibbs-sampling procedure by moving the galaxies from redshift-space to real-space according to the  peculiar motions derived from the recovered density field using linear theory with the option to include tidal field corrections from second order Lagrangian perturbation theory.
The virialised  distortions are corrected by sampling candidate real-space positions  (being in the neighbourhood of the observations along the line of sight), which are compatible with the bulk flow corrected redshift-space position  adding a random dispersion term in high density collapsed regions. The latter are defined according to the eigenvalues of the Hessian. This approach presents an alternative method to estimate the distances to galaxies using the three dimensional spatial information, and assuming isotropy. Hence the number of applications is very broad. In this work we show the potential of this method to constrain the growth rate up to $k \sim 0.3\, h$ Mpc$^{-1}$. Furthermore it could be useful to correct for photo-metric redshift errors, and to obtain improved BAO reconstructions.
\end{abstract}


\section{Introduction}

Galaxy redshift surveys produce the three-dimensional distribution of luminous sources tracing the underlying dark matter field. However, their  infered line-of-sight position is a combination 
the so-called  Hubble flow, i.e. their real distance, and their peculiar motion. The modifications produced by this are referred to as redshift space distortions (RSD). Many   astronomical studies are limited by these distortions, such as a proper environmental study \citep[][]{Nuza14}.  Nevertheless, RSD can also be used to constrain the nature of gravity and cosmological parameters  \cite[see e.g.][for recent studies]{BerNarWei01,Zha07,Jain08,Guz08,NesPer08,Song09a,Song09b,PerWhi08,McDSel09,WhiSonPer09,Song10,Zhao10,Song11}.
The measurement of RSD have in fact become a common technique \citep{CFW95,Pea01,Per04,Ang08,Oku08,Guz08,WiggleZRSD,JenBauPas11,KwaLewLin12,SamPerRac12,Reid12,OkuSelDes12,Samushia13,Zhe13,Bla13,Tor13,Samushia14,San14,Bel14,Toj14,Oku14,Beutler14}.
These studies are usually based on the large-scale anisotropic clustering displayed by the galaxy distribution in redshift-space, although $N$-body based models for fitting the data to smaller scales have been presented in \citet{Reid14}. 

In this work we use the full three dimensional distribution of galaxies and correct in a statistical Bayesian way their individual positions according to a physical model describing the relation between the density and peculiar velocity fields which depends on the growth rate. This relation  becomes more complex if one wants to correct for dispersed RSD as we will discuss below. The main finding of this work is that already very simple assumptions, such as a lognormal prior for the density field, a power-law galaxy bias, and linear theory to infer the peculiar motions with an adjustable smoothing scale, can yield accurate reconstructions of the real-space position of galaxies beyond the Kaiser-factor.

In the following section we will present our methodology and subsequently we will present the application  on a mock galaxy catalogue. Finally we will summarise our findings and present our conclusions.

\section{Method}
\label{sec:method}

Our method essentially follows the algorithm (\textsc{argo}-code) proposed in \citet{KitEns08} and \cite{KitGalFer12} by iteratively sampling the density and peculiar velocity fields within a Gibbs sampling process \citep[for the pioneering iterative correction of linear RSD see][]{Yah91}:

\ba
\mbi \delta&\curvearrowleft& {\cal P} \left(  \mbi \delta \mid N(\{\mbi r\}), \{b_{\rm p}\}, {\mat C}, \mbi w \right)\,,\\
\{\mbi r\}&\curvearrowleft& {\cal P} \left(  \{\mbi r\} \mid \{\mbi s^{\rm obs}\}, \{\mbi v(\delta,f_{\Omega},r_{\rm S},{\cal H},\{v_{\rm p}\})\} \right)\,.
\ea

In the first Gibbs-sampling step: $\mbi \delta$ is the dark matter over-density field,  ${\cal P}\left(  \delta \mid \mbi  N(\{\mbi r\}), \{b_{\rm p}\}, {\mat C}, \mbi w \right)$ the posterior distribution function of density fields given the number counts of galaxies in real-space on  a grid $\mbi N(\{\mbi r\})$, a covariance matrix $\mat C\equiv\langle\delta\delta^\dagger\rangle$ (the power-spectrum in Fourier-space), a set of  parameters describing galaxy bias $\{b_{\rm p}\}$, and a three dimensional completeness $w$.
In the second Gibbs-sampling step we obtain the real-space position $\{\mbi r\}$ from the redshift-space position of galaxies and the sampled peculiar velocity fields $\{\mbi v\}$, which depend on the large scales, on the over-density field, the growth factor $f_{\Omega}$, and a smoothing scale $r_{\rm S}$; and in the nonlinear regime additional information about the Hessian $\cal H$  and parameters describing the velocity of galaxies  $\{v_{\rm p}\}$ .

\subsection{Density field reconstruction}

We will restrict ourselves to a simple set of assumptions and demonstrate that they are sufficient for the scope set in this work, namely correct for redshift-space distortions.
 The first Gibbs-sampling iteration assumes that the galaxies are in real-space and applies a Hamiltonian sampling technique \citep{JK10},  to obtain the density field within the Lognormal approximation with stochastic bias \citep{KitJasMet10}. In particular we consider  the negative binomial distribution function with nonlinear bias, as introduced in \citet{Ata15}.  We note that the prior could be improved introducing higher order correlation functions \citep[][]{Kit12} or perturbation theory based methods \citep[see e.g.][]{Kit13}. Nevertheless, we want to remain with simple models which require the last number of assumptions and parameters.

The deterministic nonlinear bias we consider here is restricted to the following form:

\be 
\rho_{\rm g}=\gamma\rho_{\rm M}^\alpha\,,
\ee
with $\rho_{\rm h}$ being the expected number density of galaxies in a cell within a volume embedded in a  mesh, $\rho_{\rm M}$ being the dark matter density on the same mesh, $\alpha$ being the power law bias, and $\gamma$ the proportionality factor dependent on the mean number density. Here we have  neglected threshold bias, which would yield a more precise description in terms of higher order statistics \citep[][]{KYP14,Kitetal15}. We do not consider non-local bias, which will be introduced in a forthcoming publication.

\begin{figure}
\includegraphics[width=8.cm]{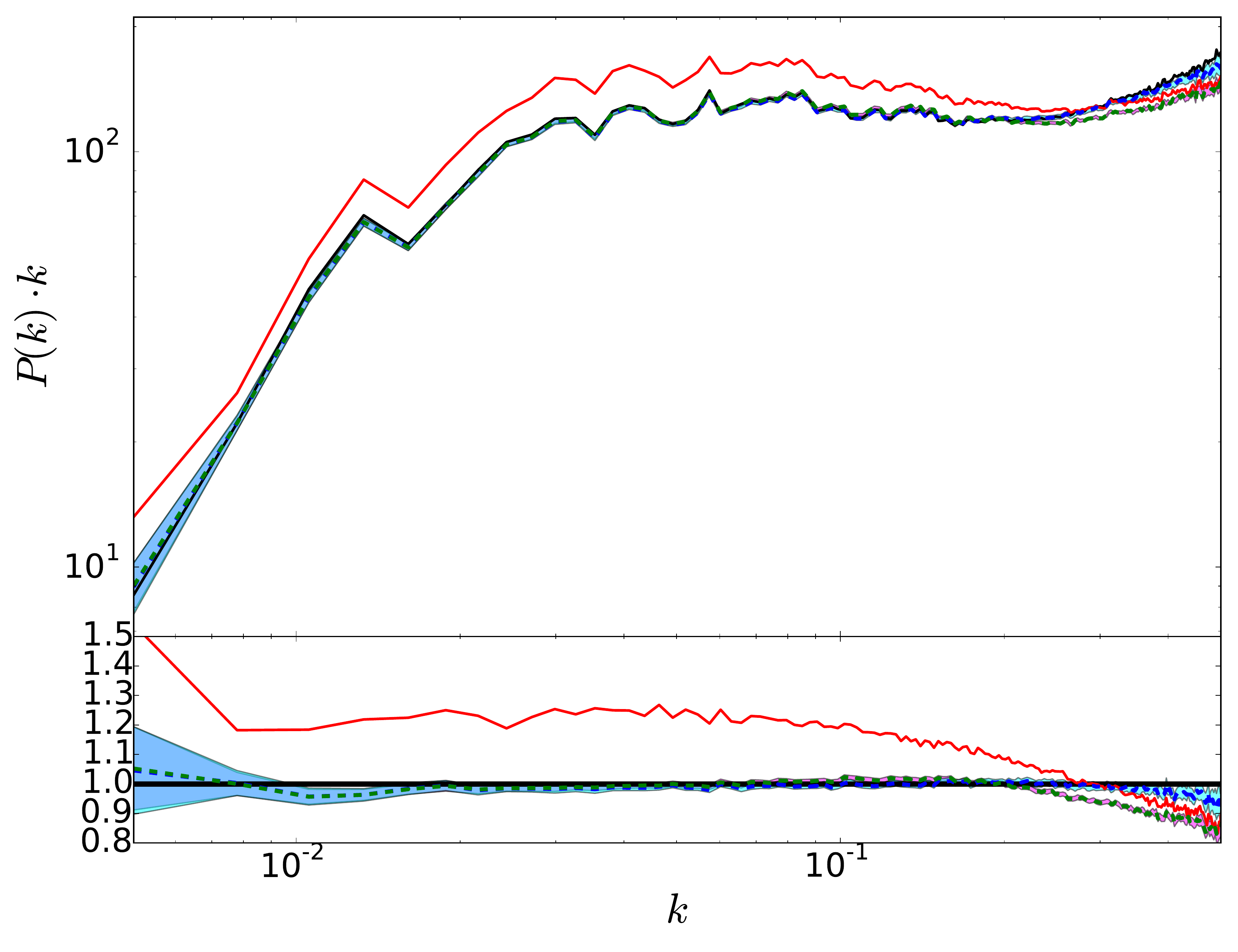}
\put(-125,-5){\rotatebox[]{0}{\text{\small [$h$ Mpc$^{-1}$]}}}
\put(-225,40){\rotatebox[]{90}{\text{\small $P^s(k)/P^r(k)$}}}
\caption{\label{fig:ps} Upper panel: power spectra of the catalog in redshift-space (red), in real-space (black); and of the reconstructed catalogues with only coherent flows  (dashed green) with 1-sigma contour based on 1000 reconstructions (magenta), coherent flows  including virialisation corrections (dashed blue) with 1-sigma contour (cyan). Lower panel: ratio between the power spectra in redshift-space $P^s(k)$ and in real-space $P^r(k)$ with the same colour code.}
\end{figure}

\begin{figure*}
\hspace{-0.5cm}
\includegraphics[width=8.cm]{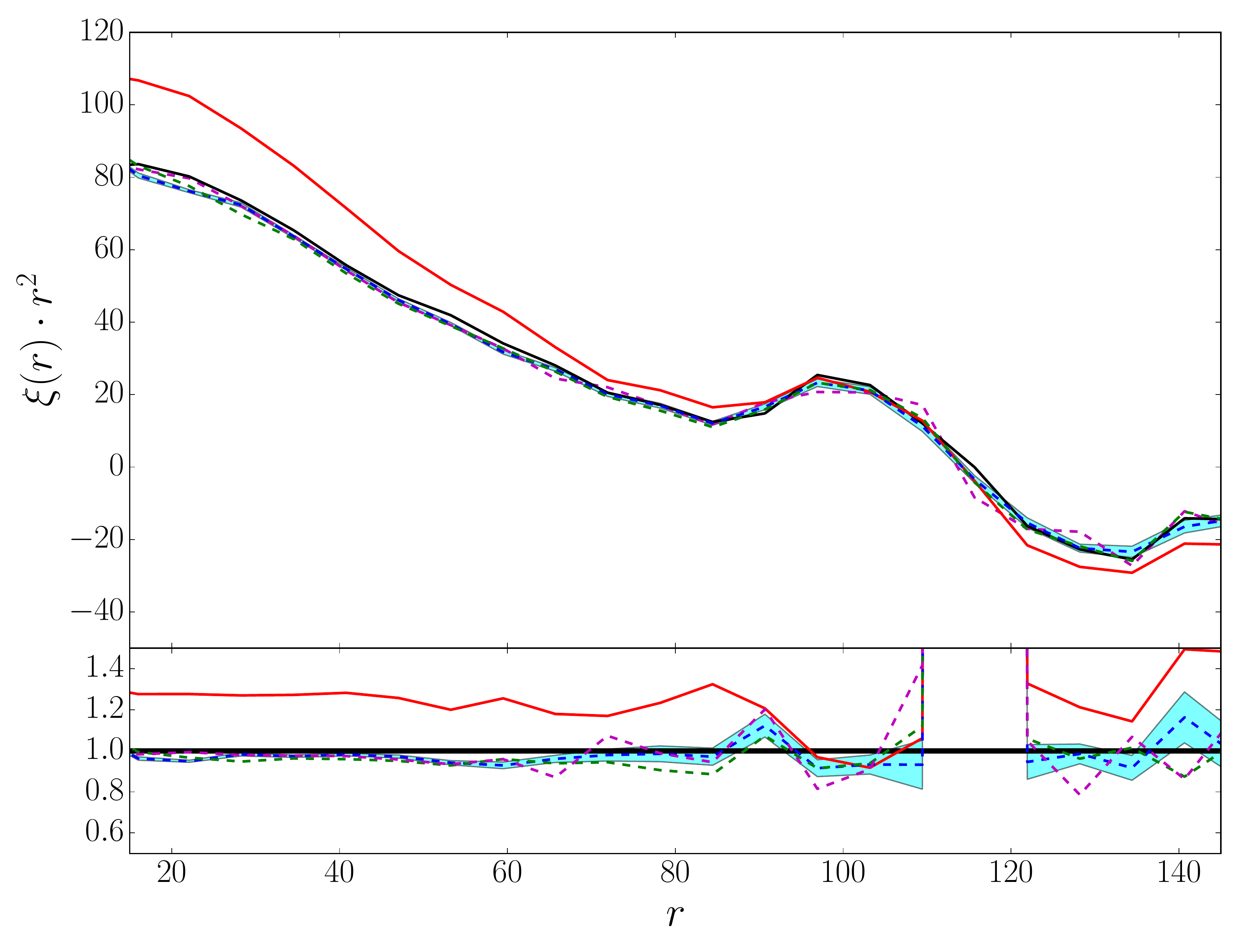}
\put(-125,-5){\rotatebox[]{0}{\text{\small [$h^{-1}$ Mpc]}}}
\put(-225,40){\rotatebox[]{90}{\text{\small $\xi^s(r)/\xi^r(r)$}}}
\hspace{1.cm}
\includegraphics[width=8.cm]{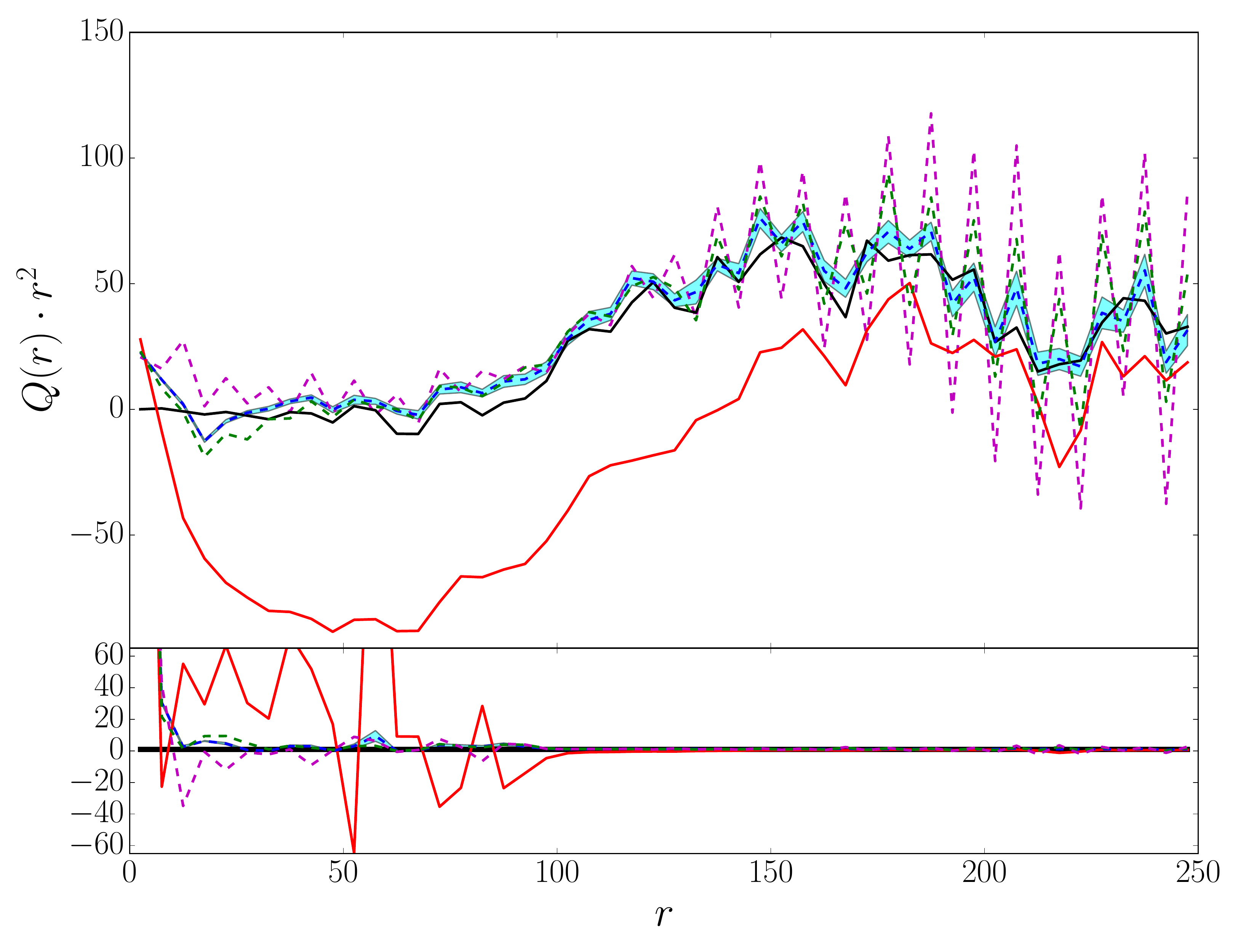}
\put(-125,-5){\rotatebox[]{0}{\text{\small [$h^{-1}$ Mpc]}}}
\put(-225,40){\rotatebox[]{90}{\text{\small $Q^s(r)/Q^r(r)$}}}
\caption{\label{fig:monoquad} Left panel: Correlation function of the catalog in redshift-space $\xi^s(r)$ (red), in real-space $\xi^r(r)$ (black); and of the reconstructed catalogues with coherent flows including virialisation corrections (dashed blue) with 1-sigma contour of 20 samples (cyan). Additionally, the reconstructed catalogues with coherent flows corrections only are shown: magenta without Gaussian smoothing, green with optimal Gaussian smoothing of $r_{\rm S}=$7 $h^{-1}$ Mpc.  Right panel: corresponding quadrupoles $Q^s(r)$, $Q^r(r)$. In the lower panels the corresponding ratios with respect to the signal in real-space are shown.}
\end{figure*}

\subsection{Peculiar motion reconstruction}
\label{sec:RSD}

The redshift-space position of a galaxy $\mbi s^{\rm obs}$ is composed by its real-space position $\mbi r$ and the coherent and dispersed components $\mbi v^{\rm coh}_r(\mbi r)$ and $\mbi v^{\sigma}_r(\mbi r)$, respectively:
\be
\mbi s^{\rm obs}=\mbi r+\mbi v^{\rm coh}_r(\mbi r)+\mbi v^{\sigma}_r(\mbi r)\,,
\ee
with the subscript r denoting the projection along the line-of-sight. Let us discuss both redshift-space contributions separately below.

\subsubsection{Coherent redshift-space distortions}

Coherent redshift-space distortions are responsible for the squashing effect of galaxy clusters along the line of sight  (see \citealp{HamiltonReview} for a review). This produces an enhancement of power on large scales, the so-called Kaiser factor \citep{Kai87}.

For simplicity we focus in this work on linear theory based on the density field ($\delta$). We are aware that improvements could be found based on the linearised density field ($\log(1+\delta)$), which we get for free in our formalism. However, it is also true that the logarithmic transformation introduces a constant offset which we do not want to consider here, as it was shown in \citet{Ney09} \citep[for more general relations including linearisations see ][and references therein]{KitAngHofGot12}. We simply consider a Gaussian smoothing (with radius $r_{\rm S}$) of the overdensity field  to optimise the velocity divergence to overdensity relation. 
For this component we do not assume any stochastic component, but directly move each galaxy from its redshift-space position $\mbi s^{\rm obs}$ to its coherent real-space position $\mbi r_{\rm coh}^{i}$ at iteration $i$, according to the coherent bulk flow motion $\mbi v^{{\rm coh},i}_r(\mbi r^{i-1})$ based on the real-space position from the previous iteration $i-1$:
\be
\mbi r_{\rm coh}^{i}=\mbi s^{\rm obs}-\mbi v^{{\rm coh},i}_r(\mbi r^{i-1})\,,
\ee
with $\mbi v_r\equiv(\mbi v\cdot\hat{\mbi r})\hat{\mbi r}/(Ha)$, where $\mbi v$ is the full three dimensional velocity field, $\hat{\mbi r}$ is the unit sight line vector, $H$ the Hubble constant and $a$ the scale factor.

\begin{figure*}
\begin{tabular}{ccc}
\hspace{0.025cm}
\includegraphics[width=4.cm]{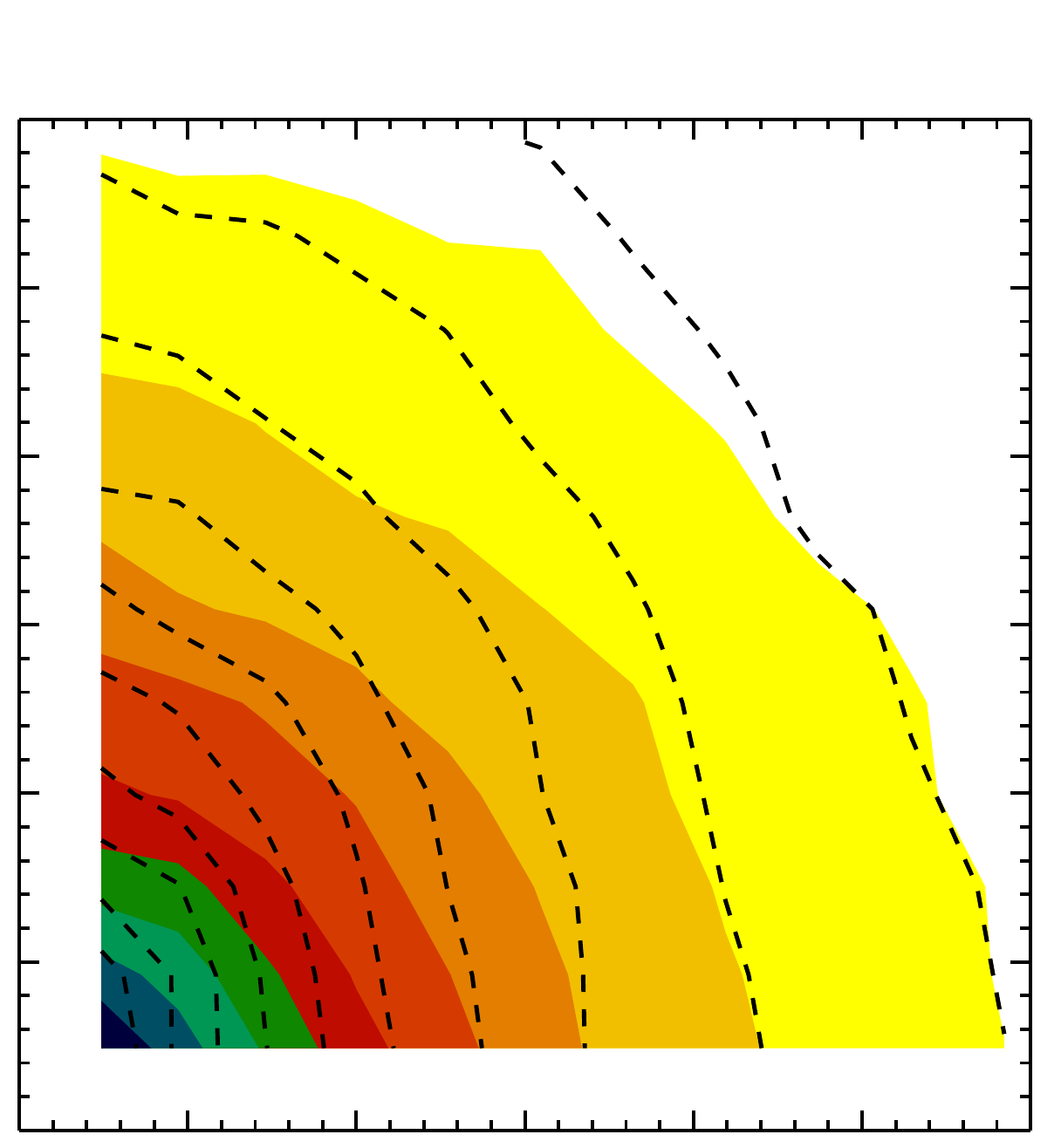}
\put(-135,55){\rotatebox[]{90}{\text{\tiny $k_{\parallel}$}}}
\put(-125,03){\text{\tiny\tiny0.00}}
\put(-125,19){\text{\tiny\tiny0.05}}
\put(-125,37){\text{\tiny\tiny0.10}}
\put(-125,55){\text{\tiny\tiny0.15}}
\put(-125,73){\text{\tiny\tiny0.20}}
\put(-125,91){\text{\tiny\tiny0.25}}
\put(-125,107){\text{\tiny\tiny0.30}}
\hspace{0.2cm}
\includegraphics[width=4.cm]{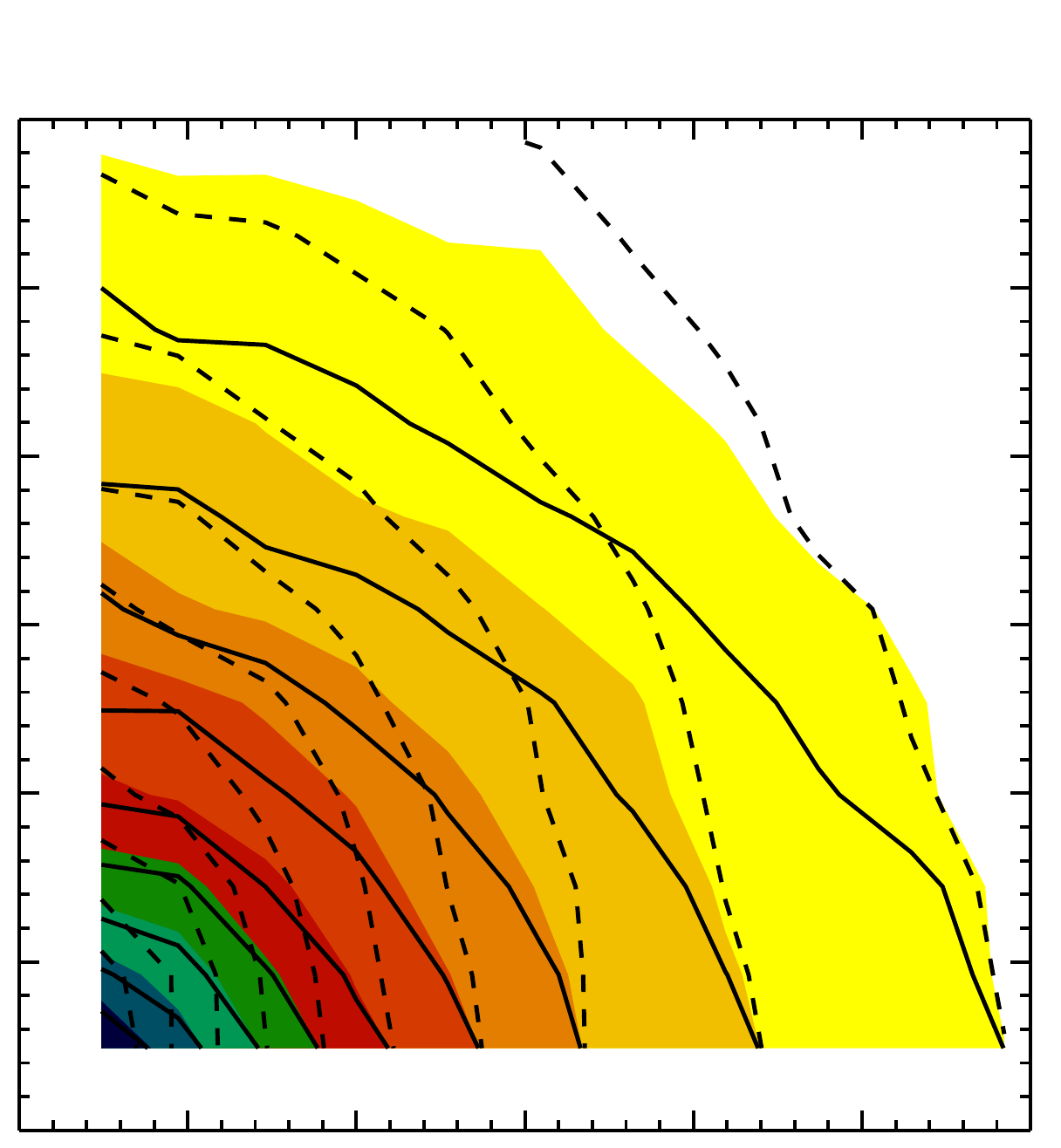}
\hspace{0.1cm}
\includegraphics[width=4.cm]{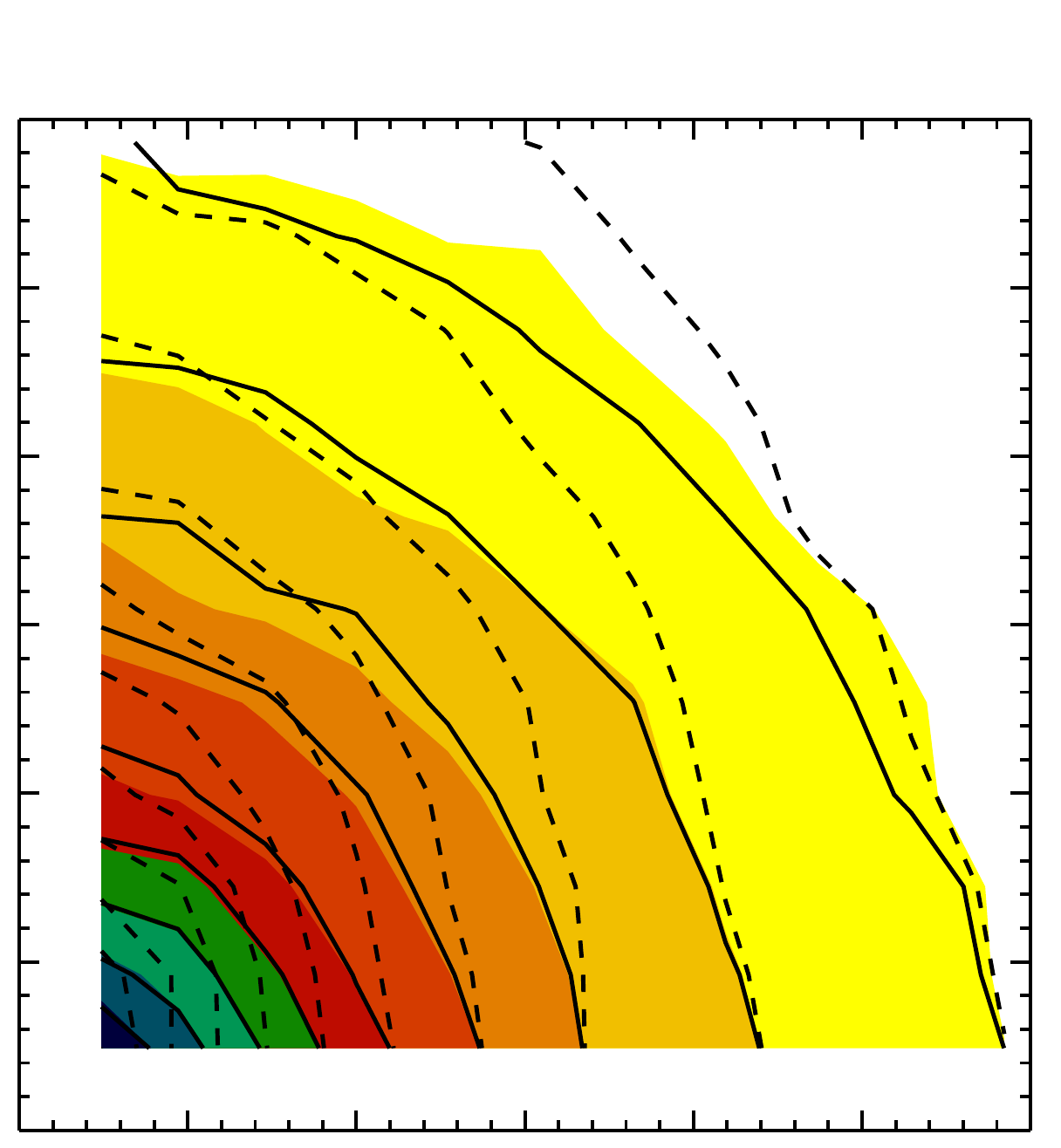}
\hspace{0.1cm}
\includegraphics[width=4.cm]{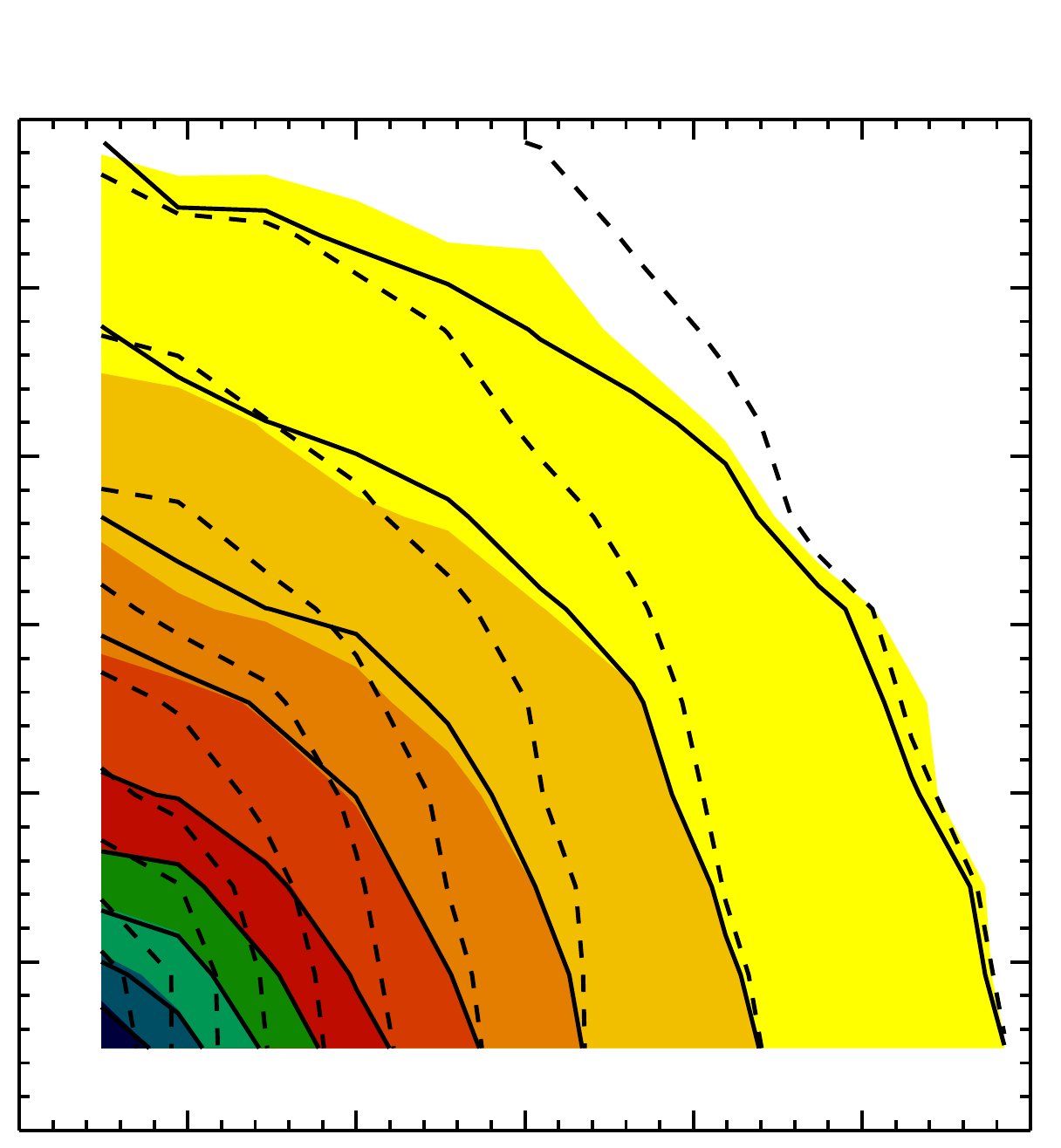}
\hspace{-0.18cm}
\includegraphics[width=.69cm]{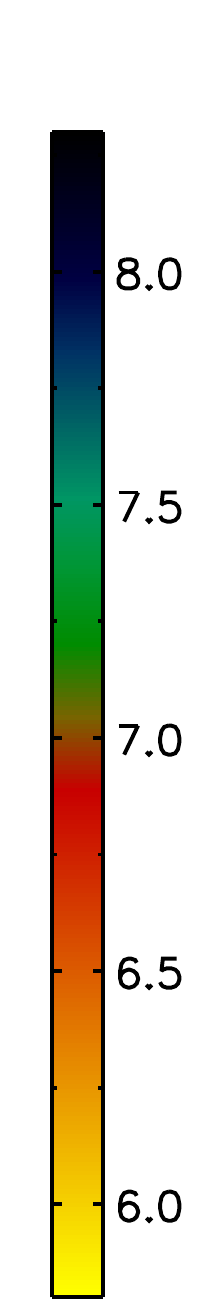}
\vspace{-.6cm}
\\
\includegraphics[width=4.cm]{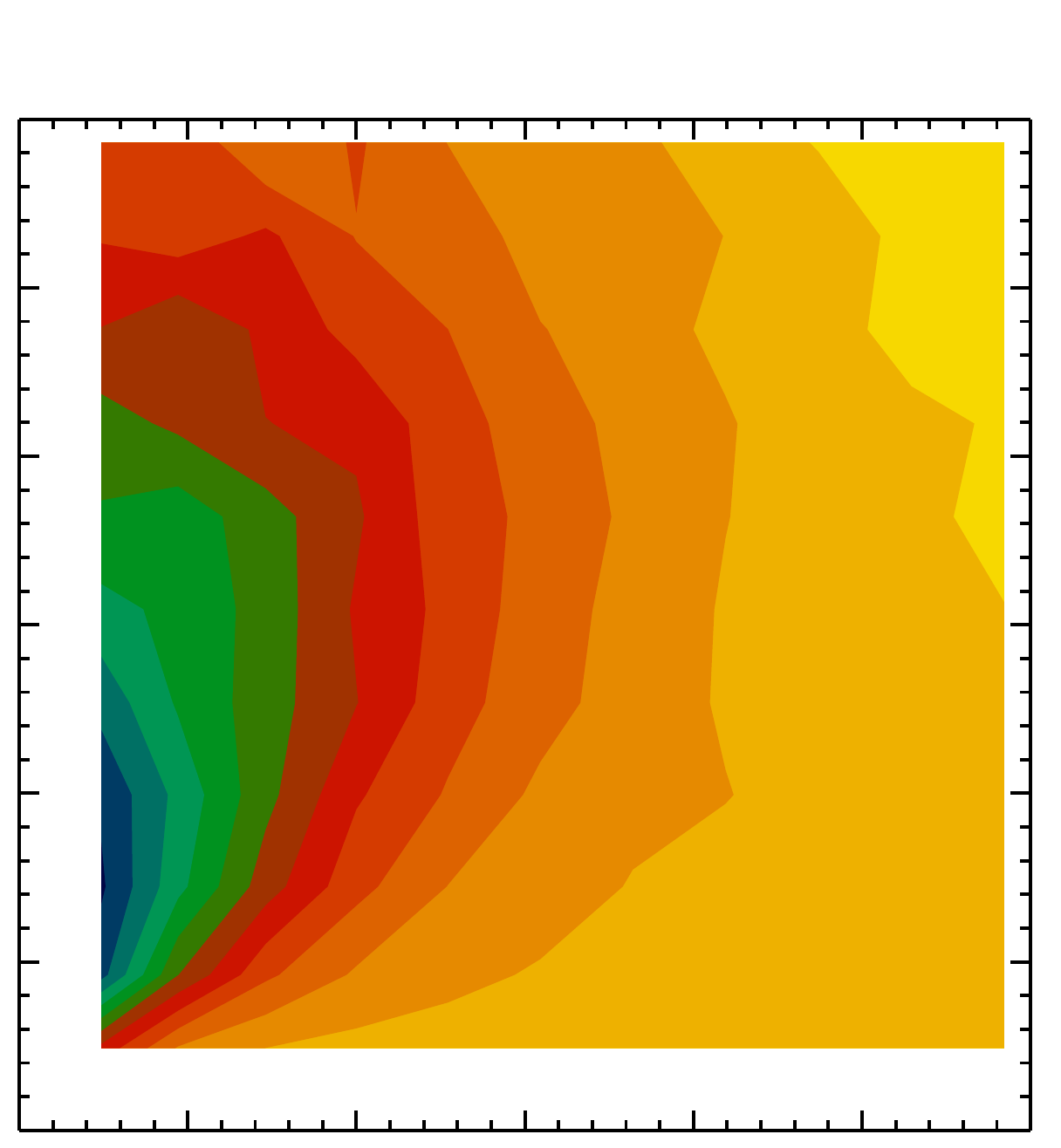}
\put(-135,115){\rotatebox[]{90}{\text{\small [$h$ Mpc$^{-1}$]}}}
\put(-135,55){\rotatebox[]{90}{\text{\tiny $k_{\parallel}$}}}
\put(-62,-10){\rotatebox[]{0}{\text{\tiny $k_{\perp}$}}}
\put(-125,03){\text{\tiny\tiny0.00}}
\put(-125,19){\text{\tiny\tiny0.05}}
\put(-125,37){\text{\tiny\tiny0.10}}
\put(-125,55){\text{\tiny\tiny0.15}}
\put(-125,73){\text{\tiny\tiny0.20}}
\put(-125,91){\text{\tiny\tiny0.25}}
\put(-125,107){\text{\tiny\tiny0.30}}

\put(-115,-3){\text{\tiny\tiny0.00}}
\put(-100,-3){\text{\tiny\tiny0.05}}
\put(-82,-3){\text{\tiny\tiny0.10}}
\put(-65,-3){\text{\tiny\tiny0.15}}
\put(-47,-3){\text{\tiny\tiny0.20}}
\put(-29,-3){\text{\tiny\tiny0.25}}
\put(-12,-3){\text{\tiny\tiny0.30}}
\hspace{0.2cm}
\includegraphics[width=4.cm]{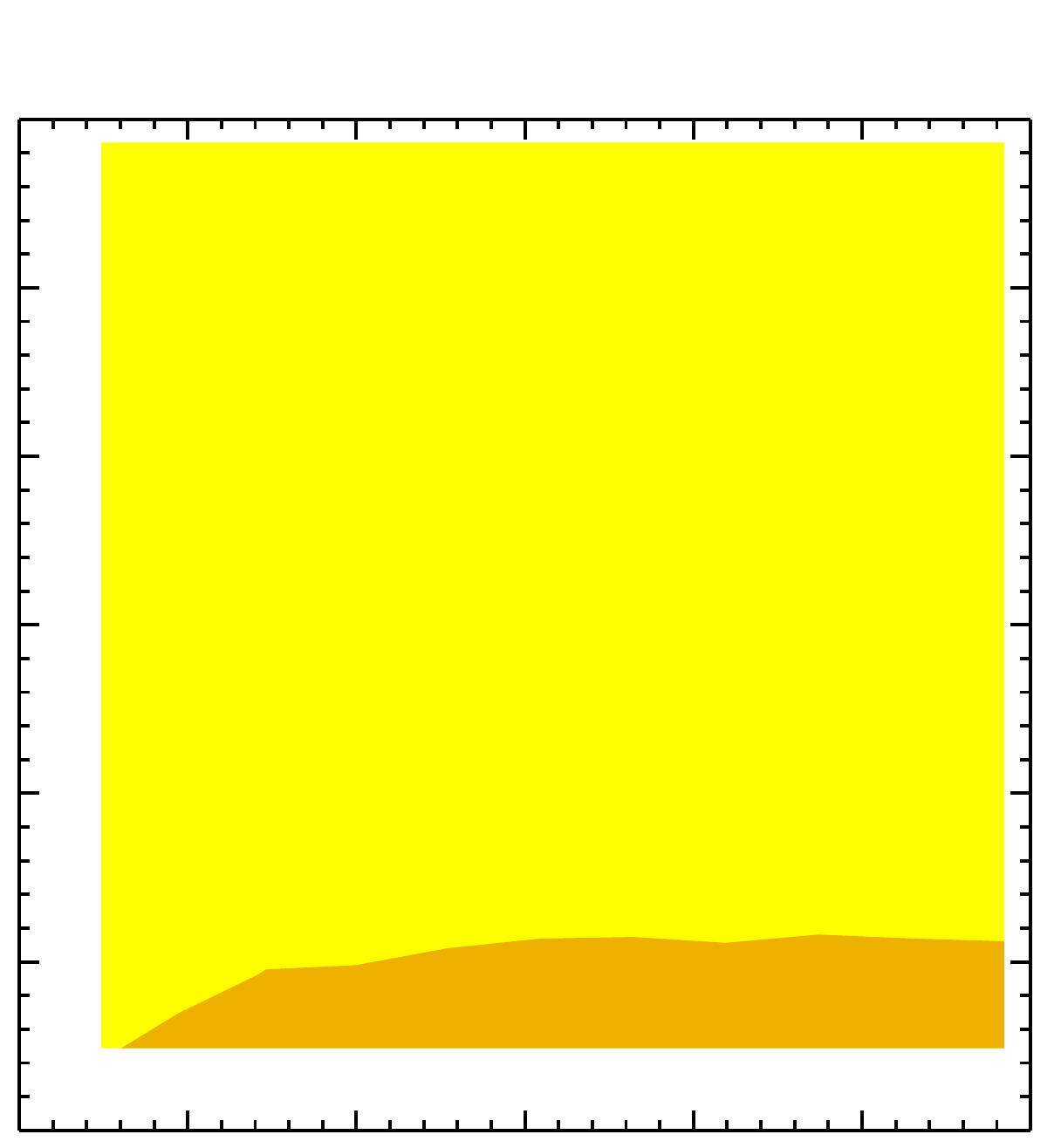}
\put(-62,-10){\rotatebox[]{0}{\text{\tiny $k_{\perp}$}}}
\put(-115,-3){\text{\tiny\tiny0.00}}
\put(-100,-3){\text{\tiny\tiny0.05}}
\put(-82,-3){\text{\tiny\tiny0.10}}
\put(-65,-3){\text{\tiny\tiny0.15}}
\put(-47,-3){\text{\tiny\tiny0.20}}
\put(-29,-3){\text{\tiny\tiny0.25}}
\put(-12,-3){\text{\tiny\tiny0.30}}
\hspace{0.2cm}
\includegraphics[width=4.cm]{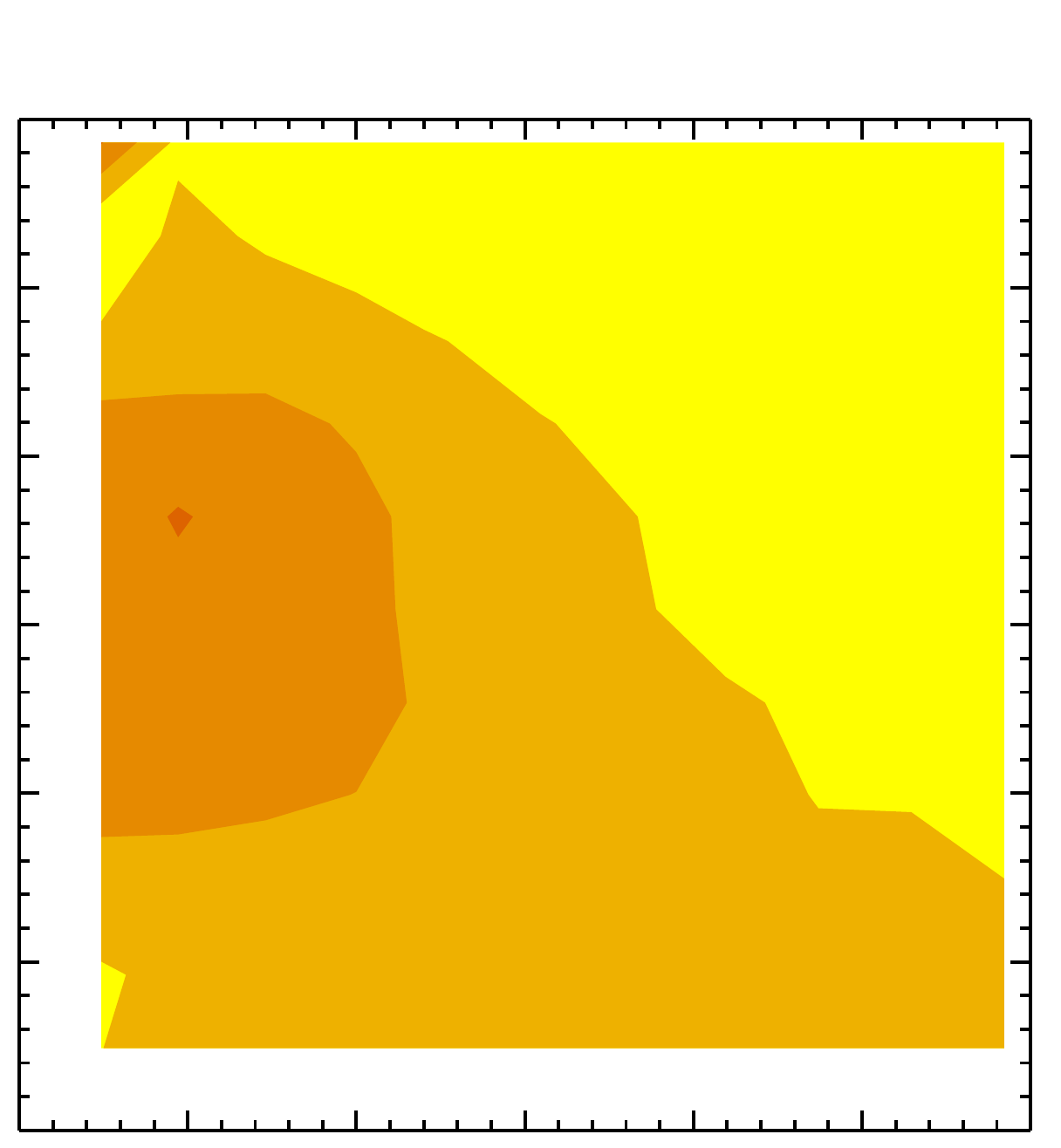}
\put(-135,-15){\rotatebox[]{0}{\text{\small [$h$ Mpc$^{-1}$]}}}
\put(-62,-10){\rotatebox[]{0}{\text{\tiny $k_{\perp}$}}}
\put(-115,-3){\text{\tiny\tiny0.00}}
\put(-100,-3){\text{\tiny\tiny0.05}}
\put(-82,-3){\text{\tiny\tiny0.10}}
\put(-65,-3){\text{\tiny\tiny0.15}}
\put(-47,-3){\text{\tiny\tiny0.20}}
\put(-29,-3){\text{\tiny\tiny0.25}}
\put(-12,-3){\text{\tiny\tiny0.30}}
\hspace{0.2cm}
\includegraphics[width=4.cm]{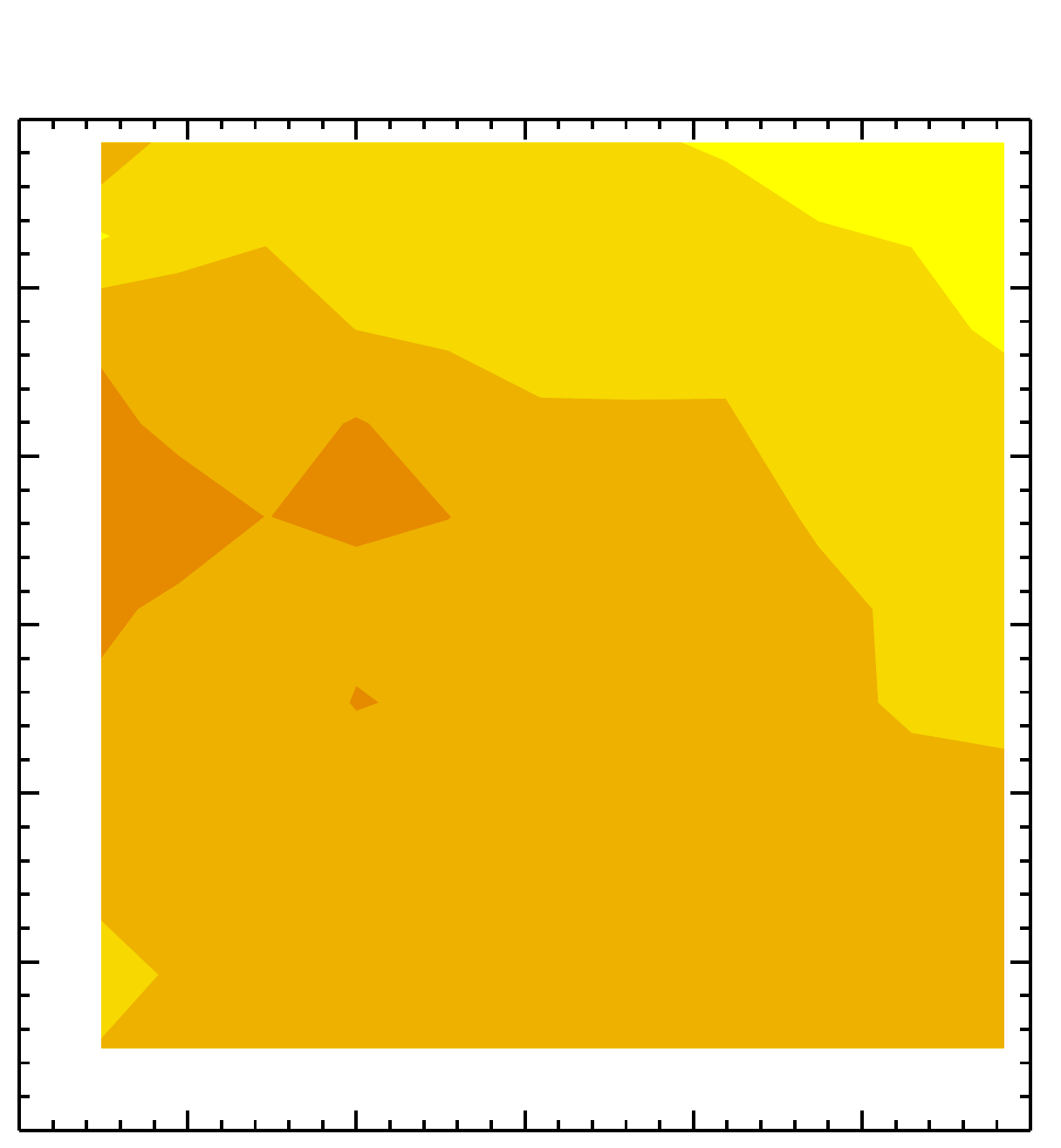}
\put(-62,-10){\rotatebox[]{0}{\text{\tiny $k_{\perp}$}}}
\put(-115,-3){\text{\tiny\tiny0.00}}
\put(-100,-3){\text{\tiny\tiny0.05}}
\put(-82,-3){\text{\tiny\tiny0.10}}
\put(-65,-3){\text{\tiny\tiny0.15}}
\put(-47,-3){\text{\tiny\tiny0.20}}
\put(-29,-3){\text{\tiny\tiny0.25}}
\put(-12,-3){\text{\tiny\tiny0.30}}
\includegraphics[width=.69cm]{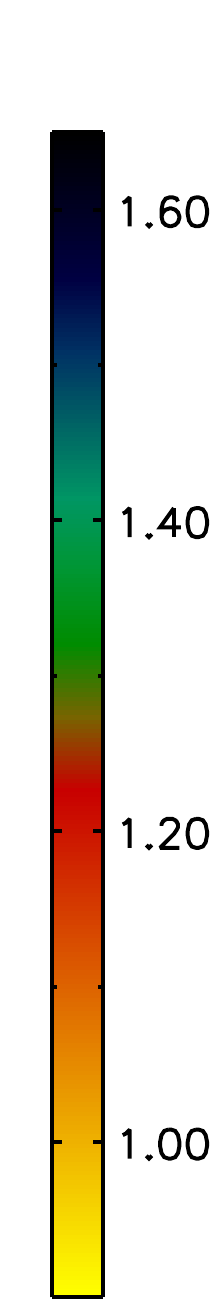}
\end{tabular}
\caption{\label{fig:2DPS}  The upper panels show the 2D power spectra $P(k_{\perp},k_{\parallel})$ corresponding to the mock galaxy catalog in real-space (colour-coded contour regions), in redshift-space (dashed lines), including the reconstructed galaxy field in real-space (solid lines): without any additional smoothing (2nd panel), with an optimal smoothing of $r_{\rm S}=7 h^{-1}$ Mpc (3rd panel),  including virialised RSD correction (4th panel). The lower panels show the corresponding ratio of the 2D power spectra, (left panel:) between the true catalogue in redshift-space and in real-space, and (for the 3 panels on the right) between the corresponding reconstructed real-space catalogue   and the true catalogue in real-space.}
\end{figure*}
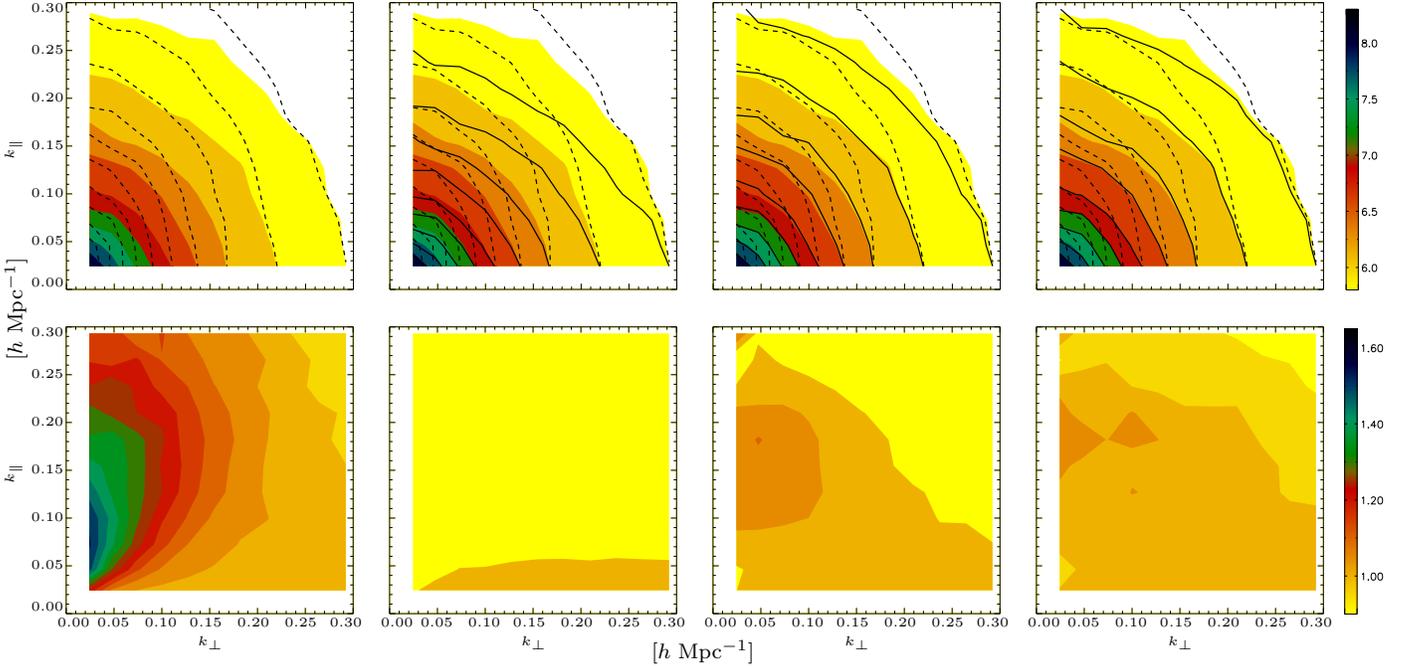

\subsubsection{Dispersed redshift-space distortions}

\label{sec:virial}

Further in the nonlinear regime in deep gravitational wells, galaxy clusters are quasi-virialised, and produced so-called fingers-of-god which 
 are elongated  structures along the line-of-sight \citep{Jac72}. This reduces the clustering towards small scales.
A method aiming at correcting these effects needs thus to enhance the clustering of galaxies in clusters. We do so by first making sure that the galaxies come from collapsed regions in real-space. To this end we demand that the real-space candidate positions along the line-of-sight come from cells in which all the eigenvalues of the Hessian are positive \citep[][]{Hahn07}  and above a certain overdensity threshold $\delta_{\rm th}$. This supposes another quantitative application of the cosmic web classification, in addition to the first of such kind presented in \citet[][]{Zhao15}.
For each real-space candidate position $\mbi r_k$ we sample the virialised motion component from a Gaussian with dispersion $\sigma(\mbi r_k)=\eta \rho_{\rm M}^\epsilon(\mbi r_k)\left(\mbi r\right)$:
\be
\mbi v^{\sigma}_r(\mbi r_k)={\cal G}\left( \mbi v^{\sigma}_r(\mbi r_k) \mid \sigma(\mbi r_k), \forall\lambda({\cal H}(\delta)>0) \right) \hat{\mbi r}\,,
\ee
with additional parameters $\eta$ and $\epsilon$ \citep[see][]{KYP14}. 

We can then find coherent real-space positions  $\mbi r_{k{\rm coh}}=\mbi r_k+\mbi v^{\sigma}_r(\mbi r_k)$ and select the closest one in each iteration:
$\min(|\mbi r_{\rm coh}-\mbi r_{k{\rm coh}}|)$. 
Additionally, to ensure that the clustering of clusters is enhanced we demand that the local density of the candidate's real-space position is larger or equal than in the previous iteration: $\delta(\mbi r_k)^i\geq\delta(\mbi r_k)^{i-1}$.
We note that a similar approach was taken using the \textsc{kigen}-code \citep{Kit13} to correct for fingers-of-god in a forward approach \citep[see][]{HKG13}. 

\section{Application to mock data}

We will focus in this work on  mock galaxies constructed using the halo abundance matching technique based on the CMASS LRG sample (see Rodr{\'i}guez-Torres et al in prep), and the BigMultiDark (\textsc{BigMD}) $N$-body simulation.

\subsection{Reference $N$-body simulation and CMASS LRG galaxy catalogue}

In particular, we performed a halo abundance matching using the redshift 0.5763 from one of the \textsc{BigMD} simulations\footnote{\url{http://www.multidark.org/MultiDark/}} \citep[][]{Klypin2014}, which was performed using the TreePM $N$-body code \textsc{gadget-2} \citep[][]{gadget} with $3840^3$ particles and the volume of $(2.5\, h^{-1}\mathrm{Gpc})^3$, in a framework of Planck $\Lambda$CDM cosmology with $\{\Omega_m = 0.307115, \Omega_b =0.048206, \sigma_8 = 0.8288, n_s = 0.96 \}$, and the Hubble parameter ($H_0 \equiv 100\,h\,\mathrm{km}\, \mathrm{s}^{-1} \mathrm{Mpc}^{-1}$) given by $h=0.6777$. We used the spherical overdensity (BDM) halo catalog. The first step to generate the galaxy catalog is to modify the maximum circular velocity ($V_{\rm max}$) of each objects adding a gaussian noise:
\be
V_{\rm max}^{\rm new}=V_{\rm max}(1+\mathcal{N}(0,\sigma))
\ee
where $\mathcal{N}(0,\sigma)$ is a gaussian random number with mean 0, and standard deviation $\sigma$. Then, we sorted all objects by $V_{\rm max}^{\rm new}$ and then, we selected objects starting from the one with larger $V_{\rm max}^{\rm new}$ and we continue until we get a number density of  3.29$\times10^{-4}h^3$/Mpc$^3$. Finally, we fit the clustering of the BOSS CMASS sample using the scatter parameter $\sigma$.  
We consider for our study the catalogue in a sub-volume of (1.25 $h^{-1}$ Gpc)$^3$.

\subsection{Reconstruction results}

 To avoid systematic effects in our study, we neglect incompleteness due to the survey geometry or the radial selection function, and neglect light-cone evolution effects. All these issues will be considered in a forthcoming work (Ata et al in prep). We compute the redshift-space position for each galaxy in the plane parallel approximation. We
then compute the corresponding density field on a cubical region with $128^3$ cells.  This permits us to make efficient computations of the 2D power spectrum. We note however, that our \textsc{argo} code has been implemented to deal with non-parallel RSD.
We obtain converged Gibbs-sampling chains after about 1000 iterations in terms of converged power spectra and also according to our convergence study demonstrated in \citep[][]{Ata15}. We have tested here different levels of deviation from Poissonity. As in \citet[][]{Ata15} we find that the stochastic bias can enhance the power towards high $k$s, however, this is not essential to our work. The smoothing scale $r_{\rm S}$ and the threshold $\delta_{\rm th}$ can be tuned to compensate for that. Since we do not aim at getting the perfectly unbiased dark matter density field in this work, but to correct for RSD, we will show only results with very high $\beta$, effectively sampling from the Poisson likelihood. An extension of this work investigating also the reconstructed dark matter field will be presented in  Ata et al. in prep.

We run three reconstruction chains, two correcting only for coherent RSDs, and the third one correcting also (partially) for virialised RSDs.

\begin{enumerate}
\item The first  RSD correction is based on coherent peculiar motions directly derived from the density field on a mesh using linear theory. 

\item The second one uses a smooth density field with Gaussian smoothing radius of 7 $h^{-1}$ Mpc obtained in a parameter study to optimally correct for RSD up to higher $k$s.  

\item The third one includes virialised corrections as described in \S \ref{sec:virial}. 
\end{enumerate}

The resulting power spectra considering 1000 reconstructions only for the latter two cases for clarity are shown in Fig.~\ref{fig:ps}. The first case yields a similar result on large scales, however underestimating the power in intermediate and small scales.
Here we can clearly see that the Kaiser factor is corrected and that the nonlinear RSD correction increases the clustering power towards small scales being closer to the true real-space catalogue.
Fig.~\ref{fig:monoquad}  shows the monopole and quadrupole in configuration space. We can see from these plots that the two different reconstructions: coherent with optimal smoothing or coherent and virialised motions corrections yield similar results. 
The left panel shows how accurately the real-space BAO can be obtained from redshift-space. However the right choice of the smoothing scale is important here, or a nonlinear virialised treatment to obtain a close BAO peak to the true one. 
We will quantify this improvement investigating also BAO reconstruction in a forthcoming work.  Interestingly, the full nonlinear RSD correction algorithm shows better agreement with the true real-space quadrupole not only on small scales being closer to zero, but also at large scales, displaying less artificial spikes, present in the pure coherent RSD corrections.
However, the different quality in the reconstructions can be better appreciated in Fig.~\ref{fig:2DPS} showing the 2D power spectra. The anisotropic pattern in redshift-space can be clearly seen. In particular, the enhancement of power due to the Kaiser effect is very prominent. A reconstruction of the peculiar velocity field with our method corrects the RSD in the catalogue in a remarkable way. We can see that an optimal choice of the smoothing scale can considerably improve the reconstruction. This is furthermore improved to scales of about $k \sim 0.3\, h$ Mpc$^{-1}$ when including virialised motions corrections.

%
%
%
%
%
%

\section{Summary and discussion}

We have presented in this work a Bayesian technique to correct for both coherent and virialised redshift-space distortions present in galaxy catalogues by estimating the distance to the individual galaxies.
We have demonstrated that this technique is accurate at least up to $k \sim 0.3\, h$ Mpc$^{-1}$ in the isotropisation of the 2D power spectrum based on precise galaxy mock catalogues describing the CMASS LRG sample. Although the method is general enough there to be precise down to far smaller scales, as indicated by the recovered power spectra. 

An application of this technique to the BOSS DR12 data including power spectrum sampling will be presented in a subsequent publication (Ata et al. in prep).

While traditional RSD measurements focus on the growth rate, an approach like the one presented in this work is complementary and more general. We refer to a recent application of a joint analysis of density fields and anisotropic power spectra including growth rate estimation, see \citet{Gra15}. The advantage of the approach presented in the present work is that it deals with nonlinear structure formation, nonlinear and stochastic galaxy bias, yielding also, as a by-product the real-space positions of the individual galaxies. 

This technique is promising for a broad number of applications, such as correcting for photo-metric redshift-space distortions including the cosmic web information, or to make precise environmental studies, as demonstrated in \citet{Nuza14} with a similar forward method recovering the corresponding primordial fluctuations. We have demonstrated in particular that it is a potentially interesting technique for the estimation of the growth rate, or for an improved BAO reconstruction. These topics will be investigated in detail in forthcoming publications using this technique.

\vspace{0.1cm}
\begin{flushleft}
{\bf Acknowledgments}
\end{flushleft}
CC, SRT, and FP acknowledge support from the Spanish MICINNs Consolider-Ingenio 2010 Programme under grant MultiDark CSD2009-00064, MINECO Centro de Excelencia Severo Ochoa Programme under grant SEV-2012-0249, and grant AYA2014-60641-C2-1-P.   GY  acknowledge support from the Spanish MINECO under research grants  AYA2012-31101, FPA2012-34694, Consolider Ingenio SyeC CSD2007-0050 and  from Comunidad de Madrid under  ASTROMADRID  project (S2009/ESP-1496). 
 FK wish to thank the Instituto de Física Teórica UAM/CSIC, and MultiDark for its support, for the hospitality during the completion of this work.
  The MultiDark Database used in this paper and the web application providing online access to it were constructed as part of the activities of the German Astrophysical Virtual Observatory as result of a collaboration between the Leibniz-Institute for Astrophysics Potsdam (AIP) and the Spanish MultiDark Consolider Project CSD2009-00064.  The  \textsc{BigMD} simulation suite have been performed in the Supermuc supercomputer at LRZ using time granted by PRACE.

\end{document}

%% file: HEADER.tex
\newcommand{\mbi}[1]{\mbox{\boldmath$#1$}}

\newcommand{\mat}[1]{\mbox{\rm\bf #1}}
\newcommand{\lsim}{\mbox{${\,\hbox{\hbox{$ < $}\kern -0.8em \lower 1.0ex\hbox{$\sim$}}\,}$}}
\newcommand{\gsim}{\mbox{${\,\hbox{\hbox{$ > $}\kern -0.8em \lower 1.0ex\hbox{$\sim$}}\,}$}}

\def\beqn{\vspace{2mm}
\begin{eqnarray}} 
\def\eeqn{\vspace{2mm} 
\end{eqnarray}}

 \def\la{\langle}

\newcommand{\be}{\begin{equation}}
\newcommand{\ee}{\end{equation}}
\newcommand{\ba}{\begin{eqnarray}}
\newcommand{\ea}{\end{eqnarray}}
\newcommand{\brr}{\begin{array}}
 
\newcommand{\err}{\end{array}}
\newcommand{\bc}{\begin{center}}
\newcommand{\ec}{\end{center}}

\newcommand{\mnras}{MNRAS}

\newcommand{\apj}{ApJ}